# A Comparative Study on the Impact of Test-Driven Development (TDD) and Behavior-Driven Development (BDD) on Enterprise Software Delivery Effectiveness.

Jun Cui[1, a, *]

1 Solbridge International School of Business, Ph.D., Daejeon, 34613, Republic of Korea.

a jcui228@student.solbridge.ac.kr

* Correspondence: Jun Cui, jcui228@student.solbridge.ac.kr.

*Abstract*—This paper compares the impact of Test-Driven Development (TDD) and Behavior-Driven Development (BDD) on software delivery effectiveness within enterprise environments. Using a qualitative research design, data were collected through in-depth interviews with developers and project managers from enterprises adopting TDD or BDD. Moreover, the findings reveal distinct effects of each model on delivery speed, software quality, and team collaboration. Specifically, TDD emphasizes early testing and iterative development, leading to enhanced code quality and fewer defects, while BDD improves cross-functional communication by focusing on behavior specifications that involve stakeholders directly. However, TDD may create a higher initial time investment, and BDD might encounter challenges in requirement clarity. These differences highlight gaps in understanding how each model aligns with varying project types and stakeholder needs, which can guide enterprises in selecting the most suitable model for their unique requirements. The study contributes to the literature by providing insights into the practical application and challenges of TDD and BDD, suggesting future research on their long-term impacts in diverse settings.

*Keywords—Test-Driven Development, Behavior-Driven Development, software delivery, enterprise effectiveness, qualitative analysis, Agile methodologies*。

## I. INTRODUCTION

In the fast-evolving landscape of enterprise software development, organizations face constant pressure to deliver high-quality software efficiently. Agile methodologies have gained traction as they promise adaptability and rapid response to changing requirements. Within this agile framework, Test-Driven Development (TDD) and Behavior-Driven Development (BDD) have emerged as popular models aimed at improving software quality and delivery speed. TDD emphasizes a test-first approach, where code is developed to pass predefined tests, ensuring functionality and minimizing errors [1-3]. BDD, by contrast, is focused on behavior specification, involving non-technical stakeholders in defining requirements through examples, which fosters shared understanding and aligns development with business needs. Despite these advantages, there is still debate over which model better supports enterprise goals for effective and timely software delivery. Although both methods are widely discussed, few studies offer direct comparisons of their impacts on delivery outcomes within enterprise settings. This research thus seeks to address this gap by examining TDD and BDD in real-world enterprise contexts, offering insights into their respective strengths and limitations [3].

This study aims to answer key questions regarding the comparative impact of TDD and BDD on enterprise software delivery effectiveness. Primary research questions include:

(1) How do TDD and BDD influence software delivery speed, quality, and team collaboration within enterprises?

(2) What are the unique challenges and benefits associated with each model when applied in enterprise environments?

(3) How does each approach affect stakeholder involvement and satisfaction in project outcomes?

By addressing these questions, the study seeks to provide a nuanced understanding of the practical effects of TDD and BDD, particularly in terms of how they align with enterprise goals for agile software delivery. Exploring these questions will help clarify whether one model is generally more advantageous than the other or if certain project types benefit more from specific aspects of each approach, thus providing actionable insights for enterprises navigating these methodologies [4].

The motivation for this research stems from the increasing reliance of enterprises on agile methodologies to maintain competitiveness in software-driven industries. As digital transformation accelerates, the demand for efficient, high-quality software delivery grows, leading organizations to explore models that can meet these objectives effectively. TDD and BDD offer distinct paths to achieving agile goals, yet enterprises often struggle with choosing the most appropriate approach due to limited comparative studies on their real-world implications. While TDD provides structure and assurance of functionality through testing, BDD's collaborative focus potentially enhances communication between developers and stakeholders. However, misalignment between model capabilities and project needs can result in inefficiencies. This research thus aims to provide clear, evidence-based guidance on how TDD and BDD impact delivery efficiency, quality, and team dynamics. By examining these models' side-by-side, this study aspires to bridge existing knowledge gaps, allowing enterprises to make informed decisions about their software development strategies [2-5].

This paper is organized into five main sections to systematically explore the impact of TDD and BDD on enterprise software delivery. The **Introduction** provides an overview of the study's objectives and contextualizes the need for comparing TDD and BDD within enterprise environments. The **Literature Review** delves into existing research on TDD and BDD, highlighting their fundamental principles and reviewing studies on their benefits and challenges. In the **Methodology** section, the research design is outlined, including the qualitative approach used to gather insights from professionals experienced with TDD and BDD in enterprise settings. The **Results and Discussion** section presents key findings, organized around themes of delivery efficiency, quality, and team collaboration, and discusses these insights in light of existing literature. Finally, the **Conclusion** summarizes the study's contributions, implications for practice, limitations, and recommendations for future research. This structure ensures a comprehensive analysis, guiding readers through the rationale, execution, and outcomes of the study.

## II. LITERATURE REVIEW

**Test-Driven Development (TDD) and Its Core Principles**

Test-Driven Development (TDD) is a software development methodology that has gained prominence within agile and lean software environments, primarily due to its emphasis on early and continuous testing. Originating from Extreme Programming (XP) practices, TDD focuses on writing tests for each unit of code before the actual code itself is implemented. This test-first approach follows a structured process: developers write a small test for a feature, implement the minimum amount of code to pass the test, and then refactor the code to enhance quality and maintainability. This cycle, known as "Red-Green-Refactor," reinforces continuous testing and validation at every development stage. TDD's benefits include higher code quality, reduced bug rates, and greater confidence in code functionality as it matures. Additionally, it facilitates clear, modular code that is easier to maintain and extend. Despite these advantages, TDD has its limitations, particularly in its heavy reliance on unit tests, which may not capture complex, system-level issues. It also requires developers to adopt a disciplined mindset, as the process can be time-consuming initially, especially in projects with evolving requirements (see **Figure 1**). Moreover, TDD may present challenges in collaborative settings if not all team members are skilled in writing tests. While TDD is highly effective for developers seeking rigor and structure, it may lack the flexibility and stakeholder involvement that other methodologies, such as BDD, offer [3-8].

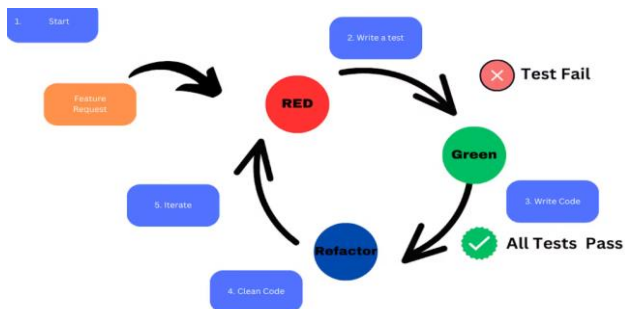

**Figure 1.** TDD Process of Software R& D.

**Behavior-Driven Development (BDD) and Stakeholder Engagement**

Behavior-Driven Development (BDD) extends beyond the developer-focused approach of TDD by incorporating a broader, behavior-oriented perspective. BDD emerged to address limitations in traditional test-driven processes by emphasizing collaboration and shared understanding between technical and non-technical stakeholders. In BDD, requirements are described in a format that defines the expected behavior of software, typically through user stories or scenarios written in natural language. This format ensures that all stakeholders, from business analysts to developers, have a common understanding of what the software is intended to do, bridging the gap between business requirements and technical implementation. BDD follows a "Given-When-Then" structure to outline scenarios, which guides developers in implementing functionality based on specific user needs and behavior. Expected benefits of BDD include enhanced communication among team members, clearer alignment with business goals, and improved stakeholder satisfaction. However, BDD also presents certain challenges, especially in defining precise scenarios, as requirements can sometimes be too broad or ambiguous, leading to implementation issues. Moreover, while BDD fosters a collaborative environment, it requires consistent stakeholder involvement, which can be difficult in enterprise settings with time constraints or when domain knowledge is limited (see **Figure 2**). Nevertheless, BDD's focus on stakeholder engagement provides valuable context that can improve software quality, though it may sacrifice the detailed, test-oriented rigor associated with TDD [5-8].

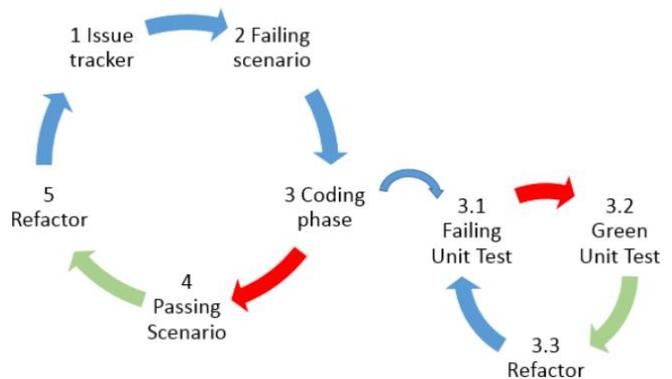

**Figure 2**. BDD Process of Software R& D.

**Comparative Analysis of TDD and BDD in Enterprise Software Delivery**

In enterprise software projects, effectiveness is often measured by key criteria such as delivery speed, quality, reliability, and alignment with business objectives. Both TDD and BDD have the potential to meet these criteria but approach them differently due to their inherent differences in process and focus. TDD, with its structured test-first development process, generally promotes high-quality code, as frequent testing minimizes defects early in the development lifecycle. However, its emphasis on unit tests and technical precision can result in a lack of alignment with broader business goals, as it doesn't involve stakeholders directly. BDD, on the other hand, offers a framework for aligning development with business expectations by involving stakeholders in the requirements definition phase. This involvement can enhance delivery reliability and reduce the risk of rework by ensuring that developers are clear on expected behavior from the start. Yet, BDD may trade off some of the rigor in testing for broader collaboration, which can impact the level of detail in testing processes. Previous studies comparing TDD and BDD indicate that both models contribute positively to software quality and delivery effectiveness, but there are gaps in understanding their unique impacts on different types of projects. For instance, while TDD may be advantageous for technical projects requiring high code quality and reliability, BDD's collaborative approach may better serve projects needing frequent stakeholder input. The theoretical frameworks of Agile and Lean Software Development support both TDD and BDD in promoting flexibility and responsiveness, yet they also highlight that the success of these models often depends on project context and team composition. This study aims to address gaps in the literature by exploring how TDD and BDD impact enterprise software delivery under varied conditions, offering insights into their respective benefits, limitations, and suitability for different project requirements [7-9].

The theoretical foundations of Test-Driven Development (TDD) and Behavior-Driven Development (BDD) are deeply rooted in Agile and Lean Software Development principles, which emphasize adaptability, iterative progress, and collaborative alignment with business goals. TDD, originating from Extreme Programming (XP), applies a structured "Red-Green-Refactor" approach, which not only enhances code quality through early testing but also enforces a disciplined, incremental development process. This systematic test-first model resonates with Agile's commitment to deliver high-quality, functional software in a flexible environment, though it focuses more narrowly on technical accuracy than stakeholder engagement. Conversely, BDD, developed as a response to the limitations of TDD, emphasizes collaboration by framing requirements through natural language scenarios, making them accessible to both technical and non-technical stakeholders. Using the "Given-When-Then" format, BDD clarifies the expected behavior of software, aligning development outcomes with user and business expectations. This stakeholder-centric model reflects Lean principles by reducing wasteful rework and enhancing shared understanding, which is critical for complex projects involving cross-functional teams. Though both TDD and BDD align with Amile's core values of iterative improvement and rapid response to change, they differ in their practical focus: TDD prioritizes code-level reliability, while BDD aims to bridge the communication gap between development and business perspectives. This study, therefore, leverages these theoretical frameworks to investigate how TDD's rigor and BDD's stakeholder engagement impact software delivery effectiveness in enterprise contexts, providing insights into their respective strengths and limitations [9-10].

III. METHODS AND MATERIALS

**Research Design**

This study adopts a qualitative case study design with a comparative analysis approach to examine the effects of Test-Driven Development (TDD) and Behavior-Driven Development (BDD) on software delivery effectiveness in enterprise environments. Given the study's focus on in-depth understanding, qualitative methods are well-suited to capture nuanced insights into how these development models influence real-world practices. By using a case study framework, the research can provide detailed perspectives on the experiences of developers and stakeholders working within TDD or BDD environments. This design allows for an exploration of factors such as development speed, code quality, and team collaboration that might vary across the two models. Comparative analysis between TDD and BDD is essential here, as it enables the identification of model-specific strengths and limitations, ultimately revealing patterns that can inform best practices. The qualitative approach thus supports a rich, contextualized analysis, making it possible to discern the specific impacts of each model on enterprise software delivery and offering valuable insights into their practical applications. (see **Figure 1**).

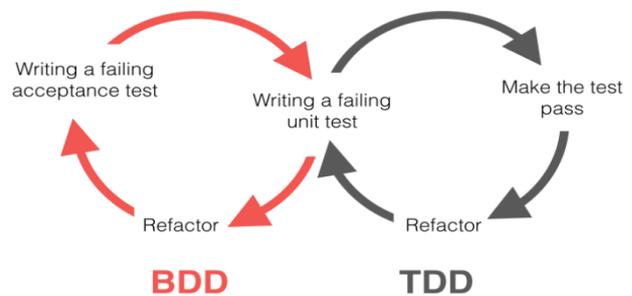

**Figure 1.** BDD and TDD cycle process.

**Data Collection**

Data were gathered primarily through semi-structured interviews with developers, project managers, and other relevant stakeholders within enterprises that use either TDD or BDD. The semi-structured format provided flexibility, allowing participants to elaborate on their experiences while also enabling the researcher to ask follow-up questions on specific aspects of TDD and BDD. Questions were designed to probe into areas such as development efficiency, code quality, collaboration practices, and challenges encountered with each model. This approach ensured that both broad and model-specific insights were collected, revealing not only the operational impacts of TDD and BDD but also the interpersonal and organizational dynamics influenced by these models. In addition to interviews, supplementary documents such as project reports and guidelines (where available) were reviewed to contextualize participants' responses. This multi-faceted data collection strategy helped to ensure a comprehensive understanding of how TDD and BDD function within organizational settings, capturing a holistic view of their contributions to software delivery effectiveness [10-13].

**Sampling**

The study used a purposive sampling approach, selecting participants based on their experience with either TDD or BDD and their involvement in relevant projects. Criteria for participant selection included a minimum of three years of hands-on experience with either TDD or BDD, as well as participation in projects of substantial size or complexity, allowing for a better understanding of how each model performs under varying conditions. Participants were drawn from different roles within the development team, including developers, project managers, and product owners, to ensure a diverse range of perspectives on TDD and BDD practices. This sampling strategy aimed to capture insights across different levels of involvement in the development process, from technical execution to project oversight, enriching the analysis with multiple viewpoints. By focusing on experienced professionals involved in real-world projects, the study ensured that the findings were grounded in practical experience, providing reliable insights into how TDD and BDD impact software delivery within enterprise settings [10-13].

### Data Analysis

The data collected through interviews and document review were analyzed using thematic analysis, a method suitable for identifying recurring patterns and themes within qualitative data. Thematic analysis allowed for a systematic examination of responses, enabling the researcher to categorize insights related to software delivery efficiency, code quality, and team dynamics in TDD and BDD environments. Initial coding identified broad themes, which were then refined into sub-themes specific to each development model, such as test thoroughness in TDD or stakeholder communication in BDD. Comparative analysis between themes provided insights into the distinct ways each model supports or challenges enterprise software delivery goals. For example, themes related to quality assurance were further examined to understand how the early testing in TDD compares to the collaborative requirement setting in BDD. This thematic approach provided a structured framework to explore complex relationships and allowed for an in-depth comparison of TDD and BDD, helping to uncover actionable insights that could inform strategic choices in software development methodologies for enterprises [11-14].

## IV. RESULTS AND DISCUSSION

### Efficiency in Software Delivery

The analysis reveals that Test-Driven Development (TDD) and Behavior-Driven Development (BDD) have distinct impacts on the efficiency of software delivery, particularly in terms of speed and predictability. Participants noted that TDD's structured, test-first approach promotes predictability by enforcing consistent testing at each stage of development, which reduces the likelihood of encountering critical bugs late in the project [14]. This predictability, however, can come at the cost of initial development speed, as TDD requires developers to write tests before coding each feature. BDD, in contrast, was reported to enhance delivery speed in projects with high levels of stakeholder involvement. By using natural language to define requirements, BDD helps to clarify expectations early, minimizing the need for rework. However, in cases where scenarios are ambiguously defined, the BDD process can slow down as developers and stakeholders negotiate more precise requirements. Compared to the literature, these findings confirm that TDD promotes controlled progress through systematic testing, while BDD can expedite delivery when collaboration is well-facilitated but may introduce delays if requirements are vague (North, 2009).

### Software Quality and Defect Rates

In terms of software quality, TDD and BDD each bring unique strengths. TDD's rigorous focus on unit tests contributes to high code reliability and low defect rates, as developers test each component individually and address issues before they impact other parts of the code. Participants observed that TDD was particularly effective at maintaining stability in complex systems, as the granular testing approach leaves minimal room for untested functionality. Conversely, BDD focuses on behavior-driven scenarios that align with user requirements, leading to higher quality in terms of meeting end-user expectations. Stakeholders remarked that while BDD may not capture technical issues as comprehensively as TDD, it reduces the risk of functional misalignment with business goals. Compared to previous studies, which highlight TDD's emphasis on technical quality and BDD's alignment with user satisfaction (Chelimsky et al., 2010), the findings suggest that both models improve quality, albeit in different dimensions—TDD excels at code stability, while BDD enhances alignment with business requirements [14-18].

### Team Collaboration and Communication

Both TDD and BDD impact team collaboration and communication, though in contrasting ways. Participants reported that TDD's focus on technical testing fosters collaboration among developers, encouraging a shared understanding of code quality standards and best practices. However, TDD's technical language and process can sometimes limit engagement with non-developer team members, such as product owners or UX designers. BDD, on the other hand, was widely recognized for fostering cross-functional collaboration by involving team members from diverse roles in defining requirements. The use of natural language scenarios in BDD enhances clarity and inclusivity, making it easier for non-technical stakeholders to contribute. This aligns with the literature on BDD's emphasis on cross-functional communication, which can be especially valuable in agile environments (Smart, 2014). Interestingly, some participants noted that while BDD facilitates communication, it may require more effort to maintain consistent scenario definitions, especially in remote teams, where frequent communication is essential [16-18].

### Stakeholder Involvement and Satisfaction

Regarding stakeholder involvement, BDD was consistently praised for promoting higher levels of engagement and satisfaction. By framing requirements in user-friendly language, BDD makes it easier for stakeholders to understand and contribute to project requirements, which can result in a greater sense of ownership and alignment with business objectives. This contrasts with TDD, which focuses more narrowly on technical testing and is typically managed within the development team. As a result, TDD may limit stakeholder involvement, although its rigorous approach does enhance trust in the final product's reliability. The findings align with literature suggesting that BDD fosters stronger stakeholder relationships through transparent, behavior-focused requirements (Adzic, 2011). However, it was also observed that BDD's success in this area is contingent on stakeholders' availability and willingness to engage regularly, which may not always be feasible in enterprise settings. These insights suggest that while BDD enhances stakeholder satisfaction, it also requires a commitment to ongoing collaboration, which could be a challenge depending on organizational constraints [17-19].

## V. CONCLUSIONS

In conclusion, this study provides a comparative analysis of Test-Driven Development (TDD) and Behavior-Driven Development (BDD) in the context of enterprise software delivery, highlighting their unique contributions to software efficiency, quality, collaboration, and stakeholder satisfaction. The findings indicate that TDD's structured testing approach enhances predictability and technical stability, proving valuable in maintaining high code reliability. However, TDD's technical focus can sometimes limit cross-functional collaboration. In contrast, BDD facilitates greater stakeholder involvement and alignment with business goals by framing requirements in accessible language, though it may pose challenges if stakeholder engagement is inconsistent. Both models contribute distinct advantages to software development; TDD excels in defect reduction and technical rigor, while BDD strengthens cross-functional communication and user-centered quality. These insights underscore the importance of aligning development models with organizational needs and project objectives, suggesting that a hybrid approach may offer balanced benefits in complex, collaborative software environments. Future research could further explore these dynamics, particularly in remote or scaled agile settings [18-19].

**Findings**

This study has revealed important differences between Test-Driven Development (TDD) and Behavior-Driven Development (BDD) in terms of their impacts on software delivery. TDD, with its emphasis on unit testing, consistently enhances code reliability and predictability, particularly valuable in complex projects where technical stability is paramount [20]. It fosters a disciplined approach to testing, though this can sometimes slow down initial development due to the extensive test-writing process. BDD, on the other hand, strengthens alignment with end-user requirements by involving stakeholders in defining test scenarios in accessible language, which can reduce misunderstandings about project expectations. However, this process can introduce delays if scenarios are not clearly defined early. The findings suggest that both models contribute distinct strengths to software development: TDD is more effective for maintaining code quality, while BDD enhances user satisfaction by aligning functionality with business needs. These insights demonstrate that the choice between TDD and BDD should be driven by specific project requirements, team composition, and stakeholder needs.

**Contributions**

This study contributes to the growing body of knowledge on software development methodologies by offering a qualitative comparison of TDD and BDD in enterprise settings. By examining the nuances of each approach, this research provides a clearer understanding of how TDD and BDD influence different aspects of software delivery, such as code quality, efficiency, and collaboration. Unlike previous studies that have focused on these models in isolation, this study highlights their relative strengths and trade-offs, providing a more balanced perspective. These contributions are valuable for developers, project managers, and decision-makers who seek to optimize software delivery processes in alignment with their organizational goals. Additionally, the study underscores the importance of aligning methodology with team capabilities and project demands, encouraging a more nuanced adoption of TDD and BDD beyond simple technical considerations [17-20].

**Implications**

The implications of this research extend to enterprises seeking to improve software delivery effectiveness. For teams focused on technical quality and defect reduction, TDD offers substantial advantages, ensuring that code is rigorously tested at each stage. However, organizations that prioritize cross-functional collaboration and stakeholder satisfaction may benefit more from BDD, which fosters greater involvement across team roles by using accessible language for requirement definitions. This study also suggests that adopting a hybrid model, integrating aspects of both TDD and BDD, could address a wider range of project demands by combining TDD's technical rigor with BDD's emphasis on business alignment. The choice of development methodology thus becomes a strategic decision that should be carefully aligned with organizational priorities, team expertise, and stakeholder engagement levels, enabling enterprises to better manage trade-offs between quality, efficiency, and satisfaction [19-20].

**Limitations**

While this study provides valuable insights, it is limited by factors that could impact the generalizability of the findings. The sample size, though sufficient for a qualitative analysis, may not fully represent the diversity of industry practices and team dynamics across different organizational contexts. Additionally, the research focused on a specific set of industries, potentially limiting the relevance of findings to other sectors with different software delivery needs or regulatory requirements. Furthermore, since the study relied on self-reported data from participants, there is a risk of subjective bias, as individuals may interpret the benefits and challenges of TDD and BDD through personal experience. These limitations suggest that while the findings are insightful, they should be interpreted with caution, particularly when applied to broader, varied enterprise environments.

**Future Work Directions**

Future research could address these limitations by exploring the impacts of TDD and BDD using quantitative or longitudinal methods. Quantitative studies that measure metrics such as defect rates, delivery times, and stakeholder satisfaction could provide more objective data on the effectiveness of each model, enabling comparisons across larger samples. Longitudinal research, following projects over extended periods, could also offer insights into the long-term impacts of TDD and BDD on team dynamics, code quality, and project outcomes [20]. Additionally, examining hybrid approaches that integrate elements of both models could be valuable, particularly in agile and remote development environments where flexibility is key. Such studies could further inform best practices, helping enterprises make more informed decisions on methodology selection and providing deeper insights into how TDD and BDD can be optimized for evolving software delivery challenges.

ACKNOWLEDGMENT

This research has been supported/partially supported Solbridge International School of Business, Woosong university, Thanks to all contributors.

ORCID

Jun Cui https://orcid.org/0009-0002-9693-9145